# Achieving Gender Representation in Lead-Author Publications in the ASTRO3D Centre of Excellence

Kim-Vy H. Tran[1] Stuart Wyithe[1] Michelle Ding[1]

[1]ARC Centre for Excellence in All-Sky Astrophysics in 3D (ASTRO3D)

**ABSTRACT**

We examine the publication rates of 400 research-focused members in the `ASTRO3D` Centre of Excellence by gender, project, and year from January 2018 to January 2024 (six years). Of the 443 refereed publications led by `ASTRO3D` members, women were first-author on 38% which is nearly double that of the astronomy field in the same period ($\sim 20$%). We record a high-water mark in 2022: 46% of `ASTRO3D` publications were led by females and 45% of research members identified as female. Using the nine research projects in `ASTRO3D`, we show that the combination of female leadership and higher fraction of female members correlates with a higher fraction of female-led publications. We find no correlation between the fraction of female-led publications and project size. Our findings demonstrate that gender representation in refereed publications can be achieved within $\sim 5$ years by combining evidence-based recruitment strategies with representation in supervisors and collaborations. We recommend that strategies for improving STEM participation focus on both female leadership and female representation to maximize effectiveness.

# 1. Introduction

*"Astronomy is great fun... The tragedy is that thousands of women are being denied a lot of fun."* Vera Rubin (Flam, 1991)

Despite numerous efforts spanning decades, there remains a tremendous gap between the fraction of women in science compared to the broader society that supports and funds science research (National Academies of Sciences, Engineering, and Medicine, 2020). In 2021, the two largest organizations of professional astronomers report women make up only 21% (International Astronomical Union[1]) to 31% of members (Ivie & Pold, 2021). The recent Astronomy 2020 Decadal survey highlighted the continuing need to recruit women into the top leadership ranks and retaining women during academic career progression (National Academies of Sciences, Engineering, and Medicine, 2023), i.e. stemming the outflows from a "leaky pipeline" (Kewley, 2021). The continuing under-representation of women and historically marginalized groups means that our community is collectively leaving an incredible amount of potential intellectual capital, creativity, and innovation at the table.

To tackle the persistent problem of gender representation in astronomy, the `ASTRO3D` Centre of Excellence implemented a number of evidence-based strategies for recruiting and retaining a representative membership. By mid-way through the seven-year award period, the fraction of `ASTRO3D` members who identify as female reached $\sim 50$% (Kewley, Wyithe, Tran, & McCarthy, 2023) which is more than double that of the international astronomical community. `ASTRO3D` has a broad range of role models and mentors for recruiting and retaining future generations of astronomers. The gender representation spans the range of career levels and provides a pipeline of professional female astronomers who can continue onto senior roles.

Here we quantify the impact of having a gender representative community by comparing the lead-author publication rates of 400 research-focused `ASTRO3D` members (as of January 2024). Publication rates are a traditional measure of productivity and crucial factor in winning research funding, but the fraction of female-led papers remains stubbornly low. In the past decade, only about 20% of refereed astronomy publications are led by women (Böhm & Liu, 2023). We show that the astronomy community can increase the fraction of female-led publications by implementing evidence based support strategies. Our study primarily focuses on two gender groups, but we emphasize that there are broader implications for member representation of our communities.

## 2. `ASTRO3D` Membership Database

### 2.1 About the `ASTRO3D` Centre of Excellence

The `ASTRO3D` Centre of Excellence was a seven-year project funded by the Australian Government through the Australian Research Council. `ASTRO3D` unifies astronomers across nine Australian nodes and seven international partner institutions to understand the evolution of matter, light and the elements, from the Big Bang to the present day. Members join one or more of the nine main science programs and are encouraged to collaborate between programs, nodes, and partner institutions.

From the very beginning, reaching 50/50 gender representation was a key goal for `ASTRO3D` and as important as the science programs. The `ASTRO3D` Equity, Diversity, and Inclusion committee developed and coordinated resources based on existing strategies for recruiting women including from the University of Melbourne (Guillemin, Wong, & Such, 2022), workforce modeling of Australian astronomy (Kewley, 2019; Kewley, 2021), and the USA National Academies of Sciences, Engineering, and Medicine 2020 report on promising practices (National Academies of Sciences, Engineering, and Medicine, 2020). The active and visible engagement of the EDI committee is valued as an essential component of the `ASTRO3D` community, and all our EDI resources are available on the `ASTRO3D` website.

(Kewley, Wyithe, Tran, & McCarthy, 2023) report how `ASTRO3D` achieved gender parity in membership by mid-way through the seven-year award period. To summarize, `ASTRO3D` uses a top-down approach from the executive leadership to science management team to have at least 50% women. `ASTRO3D` implemented evidence-based strategies for recruitment and retention such as openly advertising for positions, having gender-balanced hiring panels, and providing leadership training spanning a range of professional roles. Retention policies and culture initiatives have been particularly effective at retaining women at rates equal to or higher than men as members progress in their academic careers.

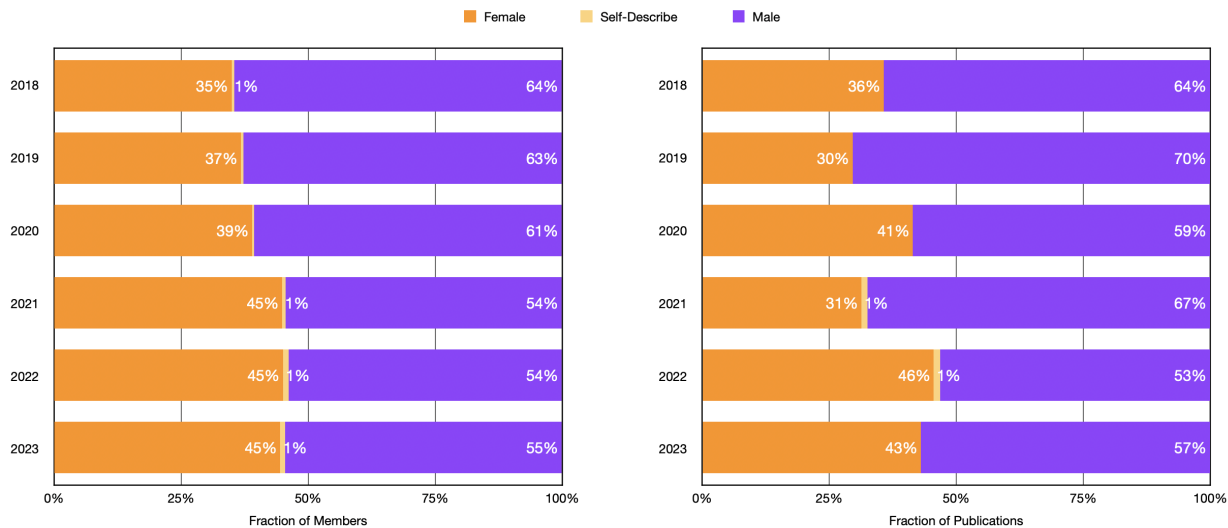

**Figure 1**

*Left:* The fraction of research-focused `ASTRO3D` members who identify as female/woman increased from 35% in 2018 to 45% by 2021. *Right:* The fraction of refereed `ASTRO3D` publications with female first-authors (orange) has increased from 36% to a high of 46% in 2022. The `ASTRO3D` average for females over the five years is 38.2% compared to 20.6% during the same period in ADS.

## 2.2 Membership Details

We use the `ASTRO3D` database that includes gender, project, and membership category for 435 members as of January 2024. As part of registering to join `ASTRO3D`, individuals voluntarily self-report their gender identity. The most common gender identities in `ASTRO3D` are female/woman and male/men (Figure 1). About 1% of responses are non-binary woman, non-binary, prefer not to say, or other. For our analysis, we combine the latter responses into a "Self-Describe" category. Because responses include female and women, we use the two identities interchangeably, and also use male and men interchangeably.

The `ASTRO3D` membership categories generally correspond to career stage (Table 1). The Students are enrolled in higher degree research programs and supervised by `ASTRO3D` members. Members with a PhD who are on fixed-term contracts include Postdoctoral Researchers and `ASTRO3D` Fellows. Members who are on continuing/long-term contracts are Associate and Chief Investigators. `ASTRO3D` also has professional staff and education/outreach focused members at the nine `ASTRO3D` nodes. Members who are not at one of the nodes are either Affiliate Researchers or Partner Investigators, i.e. they are not directly supported by `ASTRO3D` funding.

Members often change categories, especially Students and fixed-term researchers who move to other roles during the six year period of our analysis. If a member moves to a position outside of `ASTRO3D`, they become a Research Alum. In our analysis, we consider only the 400 research-focused members including Research

Alumni who publish work as part of one of the key `ASTRO3D` projects. We do not include the professional and education/outreach staff because they are not expected to publish research papers.

## 2.3 `ASTRO3D` Publications Library

Publications are a key performance indicator that is included in the official reports to the Australian Research Council and annual reports that are available to the public. The `ASTRO3D` publications library was generated using a byline search in the NASA Astrophysics Data System (ADS) for `ASTRO3D` and contains only published refereed papers. The `ASTRO3D` library is supplemented with refereed ADS publications reported by members to be `ASTRO3D` related but for which ADS missed the `ASTRO3D` byline. If members do not include the `ASTRO3D` affiliation in a publication, we assume that the research is not supported by `ASTRO3D`. The `ASTRO3D` ADS library was checked on a monthly basis by L. Staveley-Smith and is publicly available on NASA ADS.

In our analysis, we consider the 443 publications led by `ASTRO3D` members during the first six full years of `ASTRO3D` (January 2018 to January 2024; Table 2). We combine the `ASTRO3D` library with the `ASTRO3D` database to match first-author names with members and their details. In cases where members are associated with more than one project, we select the primary project. The lead-authors include active members (Students, Researchers/Fellows, Investigators) as well as Alumni who publish work completed while they were members of `ASTRO3D`.

Using the rich `ASTRO3D` database, we compare publication rates by gender to the average for the field of astronomy as a whole (Figure 2). The nine `ASTRO3D` research-focused projects vary in size ($29 - 125$ members), composition ($29 - 55$% female), and leadership (Figure 3). We examine if the proportions of publication rates by female first-authors are correlated with projects that are led/co-led by females, with fraction of females in a project, and team size (Figure 4).

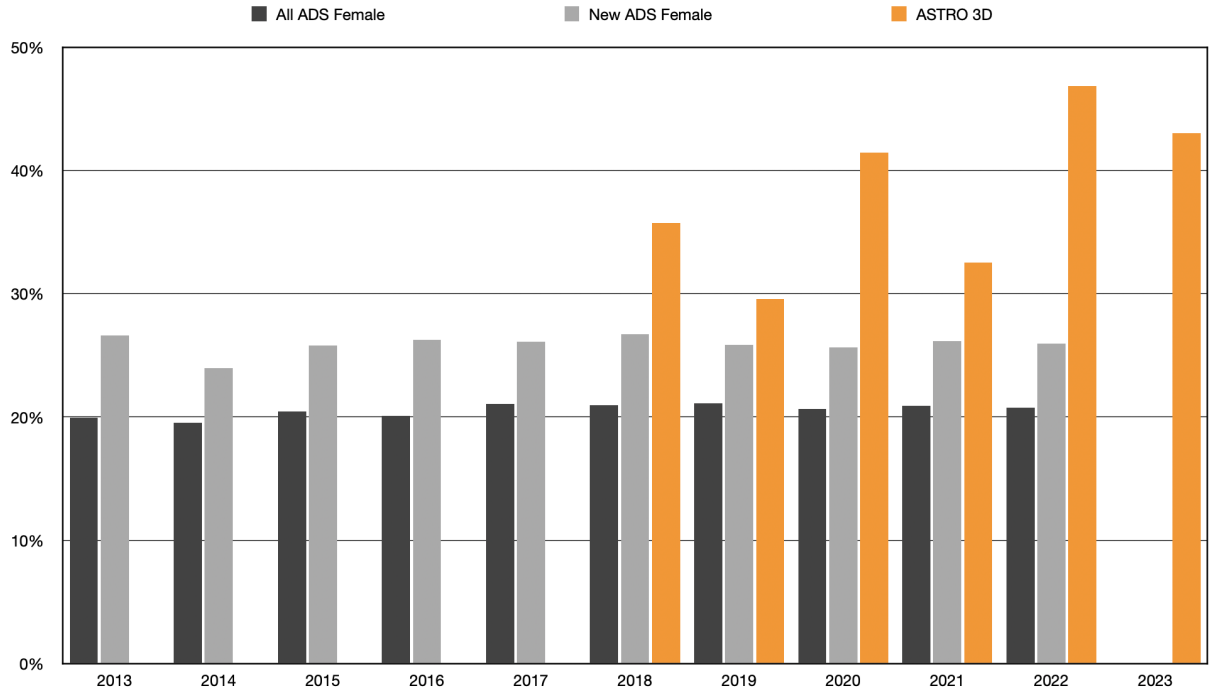

**Figure 2**

ASTRO3D has a higher average fraction of papers with female lead-authors (orange; 38.2%) compared to astronomy as a whole. The average fraction of publications with female lead authors for all publications (light gray) and for new (first-time) lead authors (dark gray) from ADS is 20.6% and 25.2% respectively and has remained stubbornly constant for the past decade . In contrast, female-led publications in ASTRO3D reached 46% in 2022, a notable achievement for any scientific organization.

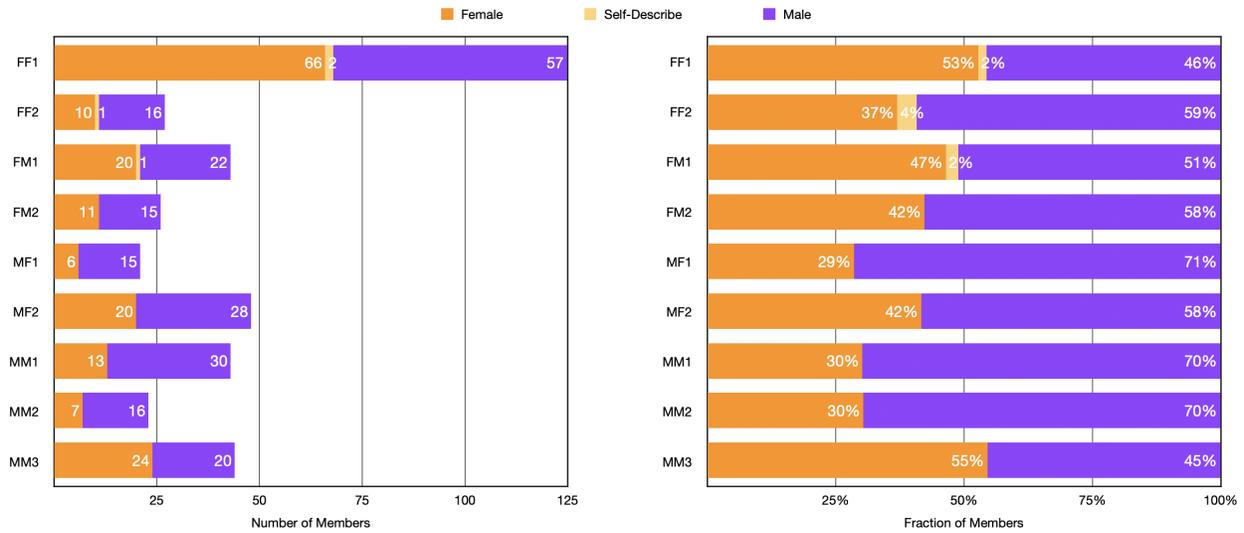

**Figure 3**
Across all of the `ASTRO3D` projects, there is a higher fraction of female members than the average of 21% for international astronomy . Each of the nine research projects has a principal lead and co-lead(s), and the projects are ordered here by project lead/co-lead where F and M stand for Female and Male. Research-focused members are assigned their primary project and include current members and research alumni. The projects span a range in size (left) and fraction of females (right).

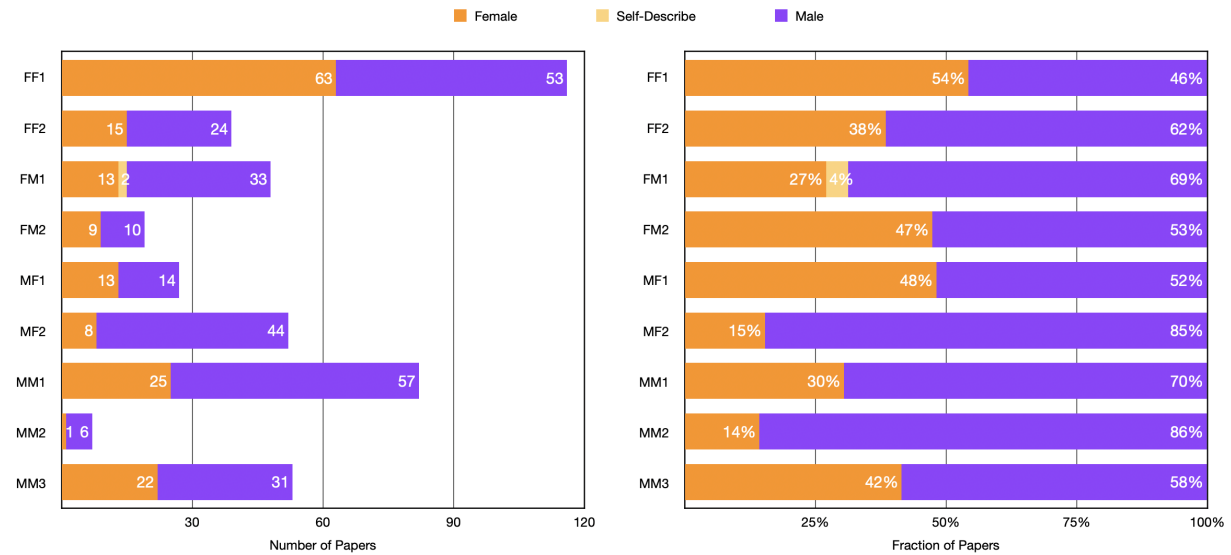

**Figure 4**
The fraction of female-led publications is correlated with projects that are led/co-led by females and with fraction of females in the project (Table 3; see Figure 3). *Left:* The FF1 project has the highest fraction of female-led publications which is consistent with the project being led by females (FF) and having one of the highest fraction of female members (53%). *Right:* Projects led/co-led by females tend to have a higher fraction of female-led publications (27-54%) compared to the projects with male leads (14-48%).

# 3. Publications

## 3.1 Compared to International Astronomy Community

`ASTRO3D` has nearly double the fraction of female-led publications compared to astronomy as a whole: 38.2% (169 of 443) of `ASTRO3D` refereed publications are led by women compared to 20.6% of all refereed publications reported in ADS during 2013-2022 (Böhm & Liu, 2023). The ADS fraction of women lead-author publications has increased by $\sim 5$% from 2005 to 2013, but the fraction has remained stubbornly low at $\sim 20$% over the past decade with no indication of increasing (Böhm & Liu, 2023). When considering women publishing their first lead author paper, the ADS fraction is higher at $\sim 25$%, but this fraction too has remained flat over the past decade (Böhm & Liu, 2023). Given the fraction of female first-author publications shows no sign of increasing from the last decade, it is highly unlikely that astronomy publications will reach ever reach parity within the existing framework.

The fraction of female-led publications within `ASTRO3D` increased from 36% to 43% from 2018 to 2023 (Table 2) and mirrors the same general increase in female members from 35% to 45% (Kewley, Wyithe, Tran, & McCarthy, 2023). A small fraction of members ($\sim 1$%) each year identify as Self-Describe, and they contribute a similar fraction of publications ($\sim 1$%) .

We note that increasing the fraction of women is likely not enough to increase the fraction of female-led publications. Even in `ASTRO3D`, having a higher fraction of women does not necessarily guarantee a higher fraction of female-led publications. The fraction of female-led publications fluctuates with lows of $\sim 30$% in 2019 and 2021, yet the fraction of female members was 37% and 45% respectively. The dip in 2021 is likely due to the gendered impact of the pandemic on research productivity (Tran, Wyithe, & Ding, 2025). The dip in 2019 is consistent within uncertainties given the fraction of women in 2019 is 37%.

## 3.2 By Research Project

`ASTRO3D` has nine research projects spanning a range in size (21-125 members) and female representation (29-55%), and each project has a principal lead and at least one co-lead (Figure 3). We test for correlations between fraction of females, project size, and project leadership using a Spearman's rank test (Table 3). We find that higher fractions of female-led publications (27-54%) correlate with female leadership. Our results are consistent with Kewley et al. (2023), who find the fraction of females in `ASTRO3D` projects that are led/co-led by females is higher than in the projects led by men.

There is also a weak correlation between fraction of female-led publications and fraction of females in the project (Figure 3 and Table 3). The FF1 project is led by females, is 53% female, and has the highest fraction of female first-author publications (54%). The MM2 project is led by men, is 30% female, and has the lowest fraction of female-led publications (14%).

Project size is not correlated with female-led publications (Table 3). Although there is a correlation between project size and a higher fraction of female members, project size alone does not result in a higher fraction of female-led publications (Table 3). Rather, we hypothesize that larger project size combined with higher female fraction and female leadership reduces the "chilly climate" (Flam, 1991) by promoting the perception of gender equity and thus fostering inclusion (Maranto & Griffin, 2011).

Our results mirror the classic study by (Kanter, 1977) who suggests that there is a tipping point when women make up $\gtrsim 35$% of the workplace. No longer a minority, women start shifting the culture and the group becomes more balanced. In a research setting with all things being equal, female leadership of and female representation in projects should produce in a higher fraction of female-led publications. A caveat of our analysis is that some of the variables are likely to be co-linear, and we cannot predict at this point the proportion of female-led publications based solely on the variables that we have measured.

## 4. Both Female Leadership and Representation Matter

We find the fraction of female first-author publications correlates most strongly with project leadership (Table 3). However, female leadership alone does not predict a higher fraction of female first-author publications (Figure 3 & Figure 4). We compare two `ASTRO3D` projects of similar size: FM1 with 43 members and MM3 with 44 members. If female leadership is the deciding factor, then FM1 should have a higher fraction of female-led publications. However, FM1 has a lower fraction of female-led publications compared to MM3 (27% to 42%).

Likewise, a higher fraction of female members does not guarantee a higher fraction of female first-author publications. Projects FM1 and FF2 are female-led with 43 and 27 members respectively, but FM1 has a higher fraction of female members (47% vs. 37%). If female fraction is the deciding factor, then FM1 should have a higher fraction of female-led publications. However, FM1 has a lower fraction of female-led publications compared to FF2 (27% vs. 38%).

Our comparisons of `ASTRO3D` projects indicate that the combination of female leadership and female representation is crucial for increasing the fraction of female-led publications. The two factors amplify the relatively weak impact that only one would have. Our results are promising but limited given the number of variables and uncertainties such as range in project sizes (21-125 members). A similar analysis across the 11 recent ARC Centres of Excellence in STEM would significantly improve the statistical robustness of our findings and further inform policies for achieving gender equity in STEM through evidence-based strategies.

## 5. Conclusions

`ASTRO3D` has achieved a remarkable milestone in gender representation regarding productivity as measured by refereed first-author publications. In the first five years, the fraction of female-led publications increased from 36% to 46% (Figure 1) and mirrors the increase in fraction of female members (Kewley, Wyithe, Tran, &

McCarthy, 2023). The average `ASTRO3D` fraction of female-led publications over the first five years was $\sim 38$%. This is in stark contrast to the field where female-led publications have hovered at $\sim 20$% for the last decade with no signs of increasing (Figure 2).

Our results indicate that strategies supporting both female leadership and female representation are more successful at achieving equity in STEM than those focusing on only one aspect. By employing evidence-based recruitment strategies, `ASTRO3D` has increased the fraction female-led refereed publications relative to the international average. Using the Spearman's rank test, we show that `ASTRO3D` projects led/co-led by females and projects with a higher fraction of females are correlated with a higher fraction of female-led publications (Table 3; Figure 3 & Figure 4). Projects with larger teams are not correlated with a higher fraction of female-led publications.

`ASTRO3D` was not immune to the impact of COVID that is evident in the low fraction of female-led publications in 2021 (Tran, Kewley, Wyithe, & Ding, 2025). However, the increase of female-led publications to $46-43$% in 2022 and 2023 indicates a resilient scientific community that is able to support all of its members. Our study demonstrates that astronomy communities can represent the broader societies that support our collective mission to explore and understand the universe.

The authors acknowledge support by the Australian Research Council Centre of Excellence for All Sky Astrophysics in 3 Dimensions ( `ASTRO3D` ), through project number CE170100013. The authors are indebted to the members who served on the `ASTRO3D` Equity, Diversity, and Inclusion Committee for their enthusiasm and engagement. Special thanks to Marie Partridge at the University of Sydney and Ingrid McCarthy at the Australian National University for their generous support of all things EDI. We thank L. Staveley-Smith for curating the ADS library and assistance with the `ASTRO3D` database and V. Böhm and J. Liu for providing their data on astronomy publications. The authors declare no competing financial interests.

| Table 1 | | | | | | |
|---|---|---|---|---|---|---|
| `ASTRO3D` Membership Categories[a] | | | | | | |
| Category | Total | Female/Woman | Male/Man | Self-Describe[b] | Node[c] | PhD Age[d] |
| Honours & Masters Students | 17 | 9 | 8 | 0 | Yes | 0 |
| PhD Students | 97 | 42 | 53 | 2 | Yes | 0 |
| Research Staff & PDRAs | 20 | 10 | 10 | 0 | Yes | $> 0$ |
| ASTRO3D Fellow | 7 | 5 | 2 | 0 | Yes | $> 0$ |
| Affiliate Researcher | 60 | 37 | 23 | 0 | No | $> 0$ |
| Associate Investigators | 65 | 22 | 41 | 2 | Yes | $\gtrsim 5$ |
| Partner Investigators | 10 | 3 | 7 | 0 | No | $\gtrsim 10$ |
| Chief Investigators | 20 | 10 | 10 | 0 | Yes | $\gtrsim 10$ |
| Research Alumni[e] | 104 | 40 | 64 | 0 | N/A | N/A |
| Professional & Education/Outreach Staff[e] | 23 | 17 | 6 | 0 | Yes | N/A |
| Non-research Alumni[e] | 12 | 10 | 2 | 0 | Yes | N/A |

| | | |
|---|---|---|
| (a) | Membership as of January 2024. We annually track individuals as they move from one research category to another, thus the number of Research Alumni increases steadily. |
| (b) | The Self-Describe group combines members who identify as non-binary woman, non-binary, prefer not to say, or other. |
| (c) | ASTRO3D has nine nodes in Australia that receive direct funding to support members. |
| (d) | Typical time since PhD awarded in years. |
| (e) | In measuring productivity with research publications, we consider only research-focused members. Professional and Education/Outreach Staff are not expected to publish astronomy papers and would bias our analysis. |

Table 2

ASTRO3D Annual Research Membership & First-Author Publications[a]

| Year | N(Fem) | N(Male) | N(SD) | NP(Fem) | NP(Male) | NP(SD) | P%(Fem) | P%(Male) | P%(SD) |
|---|---|---|---|---|---|---|---|---|---|
| 2018[b] | 59 | 109 | 1 | 15 | 27 | 0 | 35.7 | 64.3 | 0 |
| 2019 | 79 | 135 | 1 | 21 | 50 | 0 | 29.6 | 70.4 | 0 |
| 2020 | 97 | 151 | 1 | 34 | 48 | 0 | 41.5 | 58.5 | 0 |
| 2021 | 140 | 170 | 2 | 26 | 56 | 1 | 31.3 | 67.5 | 1.2 |
| 2022 | 163 | 195 | 4 | 36 | 42 | 1 | 45.6 | 53.2 | 1.3 |
| 2023 | 178 | 218 | 4 | 37 | 49 | 0 | 43.0 | 57.0 | 0 |

(a) We consider only research-focused members including alumni (see Table 1) and consider number of female, male, and self-describe: N(Fem), N(Male), and N(SD). The number of lead-author publications are listed as NP(Fem), NP(Male), and NP(SD). The corresponding fraction of lead-author publications are P%(Fem), P%(Male), and P%(SD).

(b) ASTRO3D officially started in December 2017 and had two publications led by women and one publication led by a man in 2017. We include these three 2017 publications in the 2018 values.

| Table 3 | | | |
|---|---|---|---|
| | Female Leadership, Female Fraction, and Project Size[a] | | |
| | Variable | Variable | $r_\text{Spear}$[b] |
| | Fraction Female Members | Project Size | 0.67 |
| | Fraction Female Members | Project Leads[c] | 0.37 |
| | Fraction Female-Led Publications | Project Size | 0.10 |
| | Fraction Female-Led Publications | Project Leads[c] | 0.33 |

|  | Fraction Female-Led Publications | Fraction Female Members | 0.22 |
| --- | --- | --- | --- |
| (a) | Using the nine research projects in `ASTRO3D`, we test whether the fraction of female-led publications is correlated with project size, project leadership, and female fraction in the project. | | |
| (b) | Spearman's rank correlation coefficient ranges from $-1 \leq r_{\text{Spear}} \leq 1$ where $r_{\text{Spear}} = 0$ is no correlation and $r_{\text{Spear}} = \pm 1$ is very strong correlation. | | |
| (c) | Each project has leads who are female (F), male (M), or a combination (see Figure 3). | | |

# Footnotes

1. IAU website ↩